\documentclass[12pt]{article}

\newcommand{\blind}{0}

\usepackage{latexsym, amssymb, amscd, amsthm, amsxtra, amsmath,amsthm,mathtools }
\usepackage{graphics, graphicx, color}
\usepackage{natbib}
\usepackage{ifpdf}
\usepackage{multirow}
\graphicspath{{./illustration/}}

\theoremstyle{definition}

\newtheorem*{defn*}{Definition}
\theoremstyle{remark}

\newcommand{\bi}{\begin{itemize}}
\newcommand{\ei}{\end{itemize}}
\newcommand{\be}{\begin{enumerate}}
\newcommand{\ee}{\end{enumerate}}
\newcommand{\bb}{\begin{block}}
\newcommand{\eb}{\end{block}}
\newcommand{\ba}{\begin{align}}
\newcommand{\ea}{\begin{align}}
\newcommand{\bd}{\begin{align*}}
\newcommand{\ed}{\begin{align*}}

\def\1v{\mathbf 1}
\def\0v{\mathbf 0}

\begin{document}

\def\spacingset#1{\renewcommand{\baselinestretch}%
{#1}\small\normalsize} \spacingset{1}
\if0\blind
{
  \title{\bf Combined Analysis of Amplitude and Phase Variations in Functional Data}
  \author{Sungwon Lee
    and
    Sungkyu Jung \\
    Department of Statistics, University of Pittsburgh}
  \maketitle
} \fi

\if1\blind
{
  \bigskip
  \bigskip
  \bigskip
  \begin{center}
    {\LARGE\bf Combined Analysis of Amplitude and Phase Variations in Functional Data}
\end{center}
  \medskip
} \fi

\bigskip
\begin{abstract}
When functional data manifest amplitude and phase variations, a commonly-employed framework for analyzing them is to take away the phase variation through a function alignment and then to apply standard tools to the aligned functions. A downside of this approach is that the important variations contained in the phases are completely ignored. To combine both of amplitude and phase variations, we propose a variant of principal component analysis (PCA) that captures non-linear components representing the amplitude, phase and their associations simultaneously. The proposed method, which we call functional combined PCA, is aimed to provide more efficient dimension reduction with interpretable components, in particular when the amplitudes and phases are clearly associated. We model principal components by non-linearly combining time-warping functions and aligned functions. A data-adaptive weighting procedure helps our dimension reduction to attain a maximal explaining power of observed functions. We also discuss an application of functional canonical correlation analysis in investigation of the correlation structure between the two variations. We show that for two sets of real data the proposed method provides interpretable major non-linear components, which are not typically found in the usual functional PCA.
\end{abstract}

\noindent%
\vfill

\newpage
\spacingset{1.45} 

\section{Introduction}
\label{sec:intro}

Functional data  are frequently encountered in modern sciences \citep{1}. When functional data consist of repeated measurements of a common activity or development over time, they often show a similar pattern of progression, which can be understood as a combination of two types of variations, called amplitude and phase variations. When the phase variation resides in functional data, a naive application of standard tools such as the pointwise mean and variance, and functional principal component analysis (FPCA) tends to yield misleading results \citep{3}. Curve registration (or function alignment) has been routinely performed to disregard the phase differences \citep[\emph{cf}.][]{9}. Recently, several researchers have pointed out that the phase variation also contains important information \citep{14,16,47,48,44}.

\begin{figure}[t!]
\centering
\includegraphics[width = 1\textwidth]{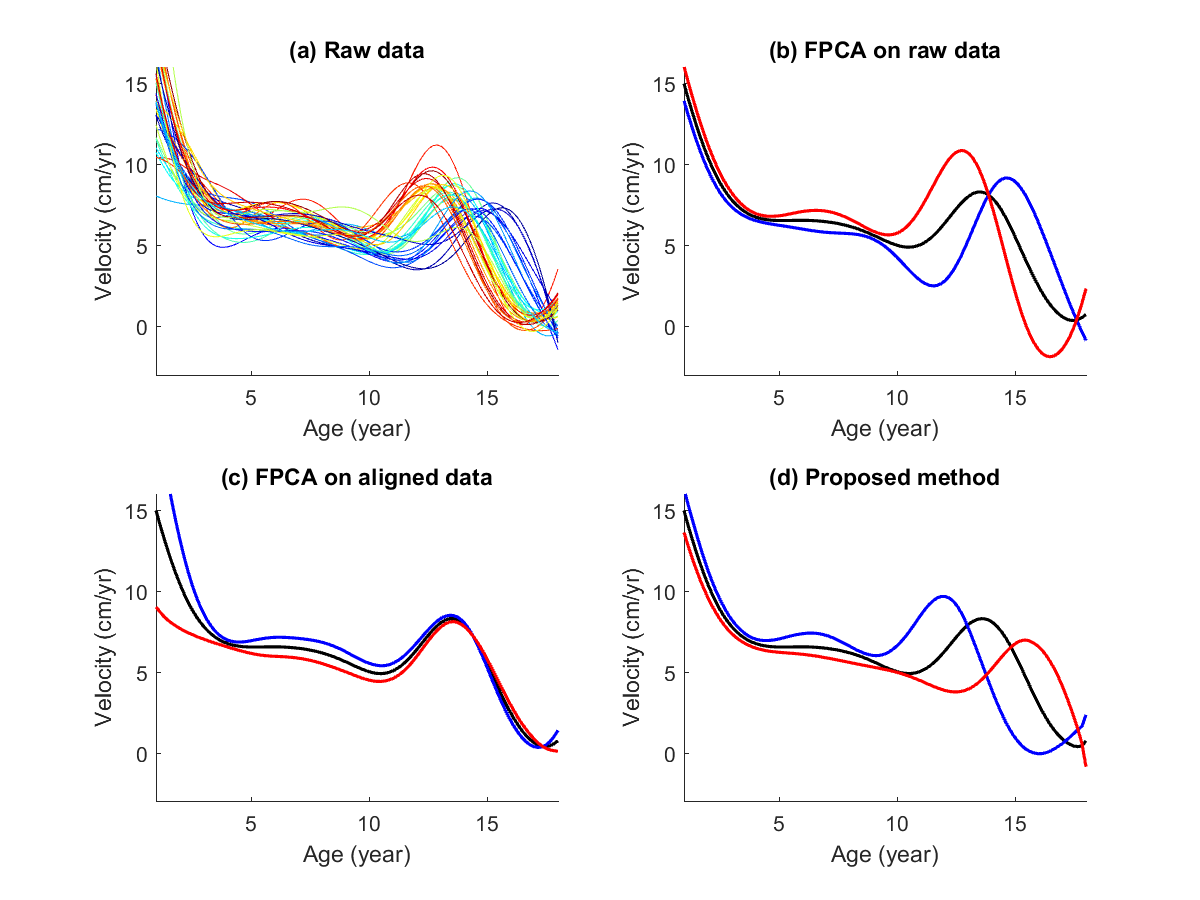}
\caption{(a) Velocity curves from the Berkeley data set, boys only.
         (b) The first component, shown as the mean (black), $\pm$2 standard deviations (red, blue) from the ordinary functional principal component analysis (FPCA). The resulting mode of variation is not easy to interpret.
         (c) The first component of the aligned data contains no information of the apparent phase variation.  (d) The first combined PC of FCPCA, the proposed method, applied to the raw data. The non-linear major variation in the data shown in (a) is well-captured by the proposed method.}
\label{fig:1}
\end{figure}

A prominent example where the phase variation is commonly observed is growth curves \citep{chen2012nonlinear,park2016clustering}. For example, the growth rate curves 
from the well-known Berkeley study \citep{6} share common events such as pubertal growth spurt and maturity. Visual inspection of this data set reveals that the curves develop the events with varying magnitudes of heights (amplitude variation), and at varying temporal paces (phase variation), as shown in in Fig.~\ref{fig:1}. 
Moreover, these two types of variations are clearly associated to each other; individuals who reach the phase of pubertal growth spurt (corresponding to the main peak of curves) later in their ages tend to show smaller maximum pubertal growth rates.
This important major association is not captured in the application of FPCA to the original data or to the aligned data (see Fig.~\ref{fig:1}(b) and (c)).

In this paper, we propose a principal component analysis for the original, unregistered data, combining the two types of variations into one. The principal components (PCs) obtained from the proposed method, which we call \emph{functional combined principal component analysis (FCPCA)}, effectively capture all of the amplitude and phase variations, including their associations. In Fig.~\ref{fig:1}, an advantage of FCPCA is exemplified for the growth data, where the dominant association between the amplitude and phase variations is well-captured in the first combined PC.

We assume that the observation $f_i$ is composed of an amplitude function,  $y_i$, and a time-warping function, $\gamma_i$, and that the observed functions can be well-aligned by time-warping functions. Our method is developed for a particular class of time-warpings, denoted by $\Gamma$, consisting of orientation-preserving diffeomorphisms of the unit interval $[0, 1]$, as developed and used in \citet{srivastava2007riemannian,13,14,16,22}.
For the combined analysis of amplitude and phase functions, we further define a bijection, denoted by $\phi$, between $\Gamma$ and a convex subset of the function space. This step enables us to use the standard linear functional operations to $x_i := \phi(\gamma_i)$.
In our FCPCA framework, we assume that the \emph{combined} random function $(y_i,x_i)$ can be represented as a linear combination of orthogonal functions. The Karhunen-Lo\`{e}ve transformation of this function is simply the FPCA in the combined function space, the components of which are then mapped back to the original function space (in which $f_i$ lies). For estimation of the functional combined components, we use a function alignment method to obtain predictions of $y_i$ and $\gamma_i$, denoted by $\hat{y}_i$ and $\hat\gamma_i$. The resulting functions $\hat{y}_i$ and $\phi(\hat\gamma_i)$ are then joined together, to which a standard functional PCA is applied.
These two functions are adaptively weighted so that the resulting combined PCs achieve the maximal explaining power of the observed functions.
The result is represented and visualized in the original function space, which can be used to aid interpretation of each principal component.

We also demonstrate a use of the functional canonical correlation analysis in the detection of maximally correlated components between the amplitude and phase functions.

In recent years, there have been a few attempts to analyze the phase variations. In particular, the phase variations are used in segmentation of periodic signals \citep{14}, clustering \citep{15}, functional regression \citep{46,47,48} and classification \citep{16}. In \citet{13,22,14,16}, the Fisher-Rao function alignment is used to obtain  time-warping functions, and the authors suggest several different approaches of analyzing the phase variation through the time-warping functions. They, however, did not discuss the association between two types of variations. While we use the Fisher-Rao alignment as used in \citet{16}, the ``composite FPCA'' of \citet{16} is less efficient than our proposal when the  amplitude and phase functions are linearly associated.
Analyses combining the phase and amplitude variations have been reported more recently in \cite{47}, where the authors used a log transformation for phase functions (thus making use of compositional data analysis). In contrast, we use the transformation $\phi$ to take advantage of the well-developed tools of conventional functional data analysis. Moreover, \cite{47} used a linear functional model consisting of individual principal component scores from each of amplitude and phase functions, which can be viewed as a two-step approach.  In contrast, we directly combine the two functions using data-adaptive weights, for the purpose of dimension reduction through non-linear principal components. Finally, \citet{chen2012nonlinear} proposed a nonparametric dimension reduction using manifold learning. Our model-based approach is conceptually different from the nonparametric approach of finding nonlinear submanifolds in  \citet{chen2012nonlinear}.

The rest of the paper is organized as follows. In Section \ref{sec:2 Models}, we formally define a population structure to model the amplitude, phase and their association, and introduce our two models, functional combined PCA and CCA. Estimation of the model parameters and the data-adaptive choice of weights are discussed in Section \ref{sec:estimation}.
In Section~\ref{sec:real data}, the advantages of the proposed methods are demonstrated in analyses of two real data sets, and in Section \ref{sec:numerical studies} several simulation studies are reported.

\section{Models} \label{sec:2 Models}

\subsection{Decomposition into two variations} \label{sec:2.1Decomp}

We consider a smooth random function $f$ that inherently contains amplitude and phase variations and is composed of two random functions $y$ and $\gamma$:
\begin{equation}
f(t) = (y \circ \gamma) (t) = y(\gamma(t)), \quad t \in [0,1].
\label{equ:1}
\end{equation}
We restrict the domain of $f$ to be $[0,1]$ without losing generality. The amplitude function $y$ is assumed to be a smooth square-integrable function on $[0,1]$, i.e.,
$ y \in L_{2}[0,1] := \{ h: [0,1] \mapsto R \mid E\| h \|_2^2  < \infty \}$.
The time-warping function $\gamma$ is an orientation-preserving diffeomorphism on $[0,1]$ and lies in
\begin{align*}
\Gamma = \{ h: [0,1] \mapsto [0,1] \mid	h(0) = 0, \,\, h(1) = 1, \,\, h^{\prime}(t) > 0, \, t \in (0,1) \} \subset L_{2}[0,1].
\end{align*}
In other words, $\Gamma$ is the set of cumulative distribution functions of absolutely continuous random variables with support on $[0,1]$.
Note that the endpoint constraints restrict the warping of $f$ to only occur on the given interval, and the positive derivative constraint does not allow the warps travel back into the past. For any $\gamma \in \Gamma$, the inverse function $\gamma^{-1}$ exists, and is also a member of $\Gamma$. This implies that $y = f \circ \gamma^{-1}$. We assume that the identity function $\gamma_{\rm id}(t) = t$ is the \emph{center} of the random warping function, where the center is defined later in Section \ref{sec:2.2 Mapping Gamma}.
This assumption formally defines phase variation as the deviation of $\gamma$ from the identity. This choice of center is purely for the sake of simplicity and interpretability; our analysis using the Fisher-Rao function alignment and the transformation of $\gamma$ discussed in Section \ref{sec:2.2 Mapping Gamma} is in fact insensitive to different choices of the center \citep[\textit{cf}. ][]{44}.

\subsection{Simplifying the geometry of $\Gamma$} \label{sec:2.2 Mapping Gamma}

Working directly with warping functions is not desirable since $\Gamma$ is not convex; there exist  $\gamma_1, \gamma_2 \in \Gamma$ and $c > 0$ such that   $\gamma_1 + \gamma_2 \notin \Gamma$ and $c\gamma \notin \Gamma$. Thus, standard operations based on Euclidean geometry can only be applied with great care. We circumvent this issue by adopting the geometric approach laid out in \citet{13,16}, and introduce a bijection $\phi: \Gamma \to B$, where $B$ is a convex subset of $L_2[0,1]$ containing the origin (i.e., the $0$ function), so that standard operations can be employed. The map $\phi$ is defined below in (\ref{eq:phifunction}), and its inverse in (\ref{equ:5b}).  The map $\phi$ is best understood as a composition of two transformations, as elaborated below.

\paragraph{Mapping to the unit sphere:} The level of difficulty in dealing with $\gamma$ is eased by taking the square-root of the derivative of $\gamma$,  the operation of which is denoted by $\Theta: \Gamma \to L_2[0,1]$,
\begin{equation}
\Theta(\gamma) := q_{\gamma} =  \sqrt{\gamma^{\prime}}.
\label{equ:2}
\end{equation}
This corresponds to the ``{sqaure-root velocity function}'' of \citet{13}.
Denote by $S_+ = \{h \in L_2[0,1] : \| h \|_2 = 1, h(t) > 0, \mbox{ for all } t \in (0,1) \}$  the positive orthant of the unit sphere in $L_2[0,1]$.
It can be checked that for any $\gamma \in \Gamma$, $q_{\gamma} \in S_+$ and that $\Theta: \Gamma \to S_+$ is a bijection.
A significant benefit of using this transformation is that the complicated structure of $\Gamma$ is simplified to that of the well-known unit sphere.

The ``center'' of the random diffeomorphism $\gamma$ is defined through the Karcher mean \citep{karcher1977riemannian} of $\Theta(\gamma)$. Let
$$\mu = \mu(\gamma) = \mathbb{E}[\Theta(\gamma)] = \mbox{arg}\min_{\mu \in S_+} E [d_g^2(\Theta(\gamma), \mu)]$$ be the Karcher mean using the geodesic distance  $d_{g}(a,b) = \cos^{-1}(\langle a,b \rangle)$. Then $\Theta^{-1}(\mu(\gamma))$ is the center of $\gamma$.

\begin{figure}[t!]
\centering
\includegraphics[width = 0.8\textwidth]{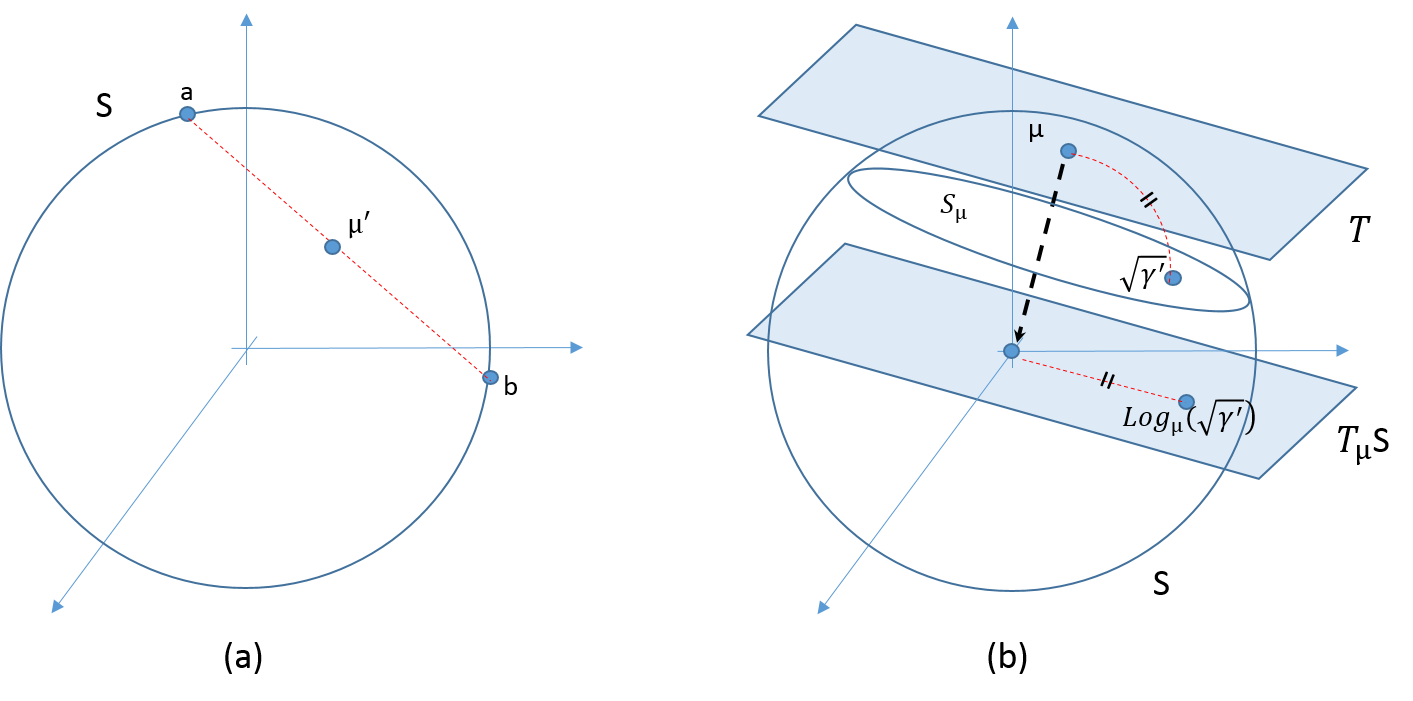}
\caption{Schematic illustration of $S$ and $T_\mu S$. (a) The pointwise mean $\mu^\prime$ of two functions $a, b \in S$ does not lie on $S$. (b) Mapping of $\sqrt{\gamma^{\prime}} \in S$ to a tangent space $T_{\mu}S$ by the log map. An example of $S_\mu$ is given as a ball in $S$ centered at $\mu$. }
\label{fig:2}
\end{figure}

\paragraph{Mapping to a tangent space:}
The positive unit sphere $S_+$ has been well-studied as a space for random directions and unit-norm random functions. While there are several approaches of modeling random elements in $S_+$ \citep[cf.][]{Mardia2000,JungPNS,16}, we use a linear approximation of $S_+$ by a tangent space. The tangent space approximation is schematically illustrated in Fig. \ref{fig:2}. For simplicity, we use the unit sphere $S$ that includes $S_+$.

The tangent space of $S$ at a point $\mu \in S$, denoted by $T_{\mu}S$, is the collection of functions in $L_2[0,1]$ orthogonal to $\mu$,
\begin{align*}
T_{\mu}S=\{h(t) \in L_2[0,1] : \langle h, \mu \rangle = 0 \},
\end{align*}
where $\langle \cdot, \cdot \rangle$ is the usual inner product in $L_2[0,1]$. Functions in $S$ will be approximated by functions in $T_{\mu}S$.
Figure \ref{fig:2}(b) schematically illustrates $T_{\mu}S$ and the approximation of the $S$-valued function $\sqrt{\gamma\prime}$ by the function $\mbox{Log}_\mu(\sqrt{\gamma\prime}) \in T_{\mu}S$; see (\ref{equ:3}). To help understand the tangent space approximation, we take the hyperplane $T$ in $L_2[0,1]$ tangent to $S$ at $\mu$. The tangent space $T_{\mu}S$ is obtained by a translation of the hyperplane $T$ so that the tangent point $\mu$ is translated to the origin. Thus, $T_{\mu}S$ is a subspace of $L_2[0,1]$.

Points (i.e., functions) on the tangent space $T_{\mu}S$ can provide good approximations of functions in a subset $S_{\mu} \subset S$ containing $\mu$. In particular, the \emph{log map} is frequently used for such approximation, and is defined as $\text{Log}_{\mu}: {S}_{\mu} \to T_{\mu} S$,
\begin{equation}
\text{Log}_{\mu}(q_\gamma) = \frac{d_{g}(q_\gamma, \mu)}{\text{sin}(d_{g}(q_\gamma, \mu))} (q_\gamma-\text{cos}(d_{g}(q_\gamma, \mu))\mu).
\label{equ:3}
\end{equation}
The geodesic distance $d_{g}(q_\gamma,\mu)$ measures the distance between $q_\gamma (=\sqrt{\gamma\prime})$ and $\mu$ by the length of the shortest arc on $S$ that joins $\sqrt{\gamma^{\prime}}$ and $\mu$. When the standard $L_2$-norm is used for $T_{\mu} S$, the geodesic distance between $\mu$ and $q_\gamma$ and the direction in which $q_\gamma$ shoots from $\mu$, for any $q_\gamma \in S_+$, are preserved by the log map.

A sensible choice of the tangential point $\mu$ is given by the assumption that $\gamma_{\rm id}$ is the center of $\gamma$. It can be seen that $\Theta(\gamma_{\rm id}(t)) = 1$ for all $t \in [0,1]$. Thus we choose the constant function $1$ as $\mu$.
This entails that the Karcher mean of $\Theta(\gamma)$ is $\mu(\gamma) \equiv 1$, and that
$E[ \log_\mu(\Theta(\gamma))] = 0$.
Note that the center, $\gamma_{\rm id} = \Theta^{-1}(1)$, of $\gamma$ is in general different from the mean of $\gamma$.

\paragraph{Summary:}
The mapping $\phi$ we use for the trasformation of the time-warping function $\gamma$ is
$\phi: S_+ \to T_\mu S$,
\begin{equation}\label{eq:phifunction}
\phi(\gamma) = \log_\mu(\Theta(\gamma)),
\end{equation}
where $\mu \equiv  1$.
We call $x = \phi(\gamma)$ phase function. Since the image of $\phi$ (denoted by $B$) is a convex subset of $T_\mu S$, standard vector operations (e.g., the Gram-Schmidt operations) can be used for the phase function $x$.

\subsection{Construction of $f$ by the amplitude and phase functions} \label{sec:2.3 Construction of f}

Any pair of amplitude and phase functions $(y,x) \in L_2[0,1] \times B$ can be composed to a single function, by reverting the decomposition in Sections \ref{sec:2.1Decomp} and \ref{sec:2.2 Mapping Gamma}. To define this composition, we note that the log map is indeed the inverse of \emph{exponential map}, $\text{Exp}_{\mu}: T_\mu S \to S$, defined by
\begin{equation}
\text{Exp}_{\mu}(x)  = \frac{\text{sin}\|x\|_2}{\|x\|_2}x+\text{cos}\|x\|_2\mu.
\label{equ:4}
\end{equation}
For any phase function $x \in B \subset T_{\mu}S$, the corresponding time-warping is uniquely given by
\begin{align}\label{equ:5b}
\gamma = \phi^{-1}(x) =  (\Theta^{-1} \circ {\rm Exp}_\mu) (x),
\end{align}
that is, $\gamma(t) =  \int_0^{t} \text{Exp}_{\mu}^2(x)(s) ds$, $t \in [0,1]$.
All in all, any random functions $(y,x) \in L_2[0,1] \times B$ can be composed to yield a random function $f$ in the form of (\ref{equ:1}) as follows.
\begin{equation}
f(t) = (y \circ \gamma) (t) = (y \circ \phi^{-1}(x) ) (t) =  y \left ( \int_0^{t} \text{Exp}_{\mu}^2(x)(s) ds \right ), \,\, t \in [0,1].
\label{equ:5}
\end{equation}

\subsection{Models for joint variability of amplitude and phase functions} \label{sec: FCPCA and FCCCA}

In this section, we define the joint population structures of the amplitude and phase functions $(y,x)$. The parameters we aim to estimate are defined in the models we describe below. Recall that the mean of $x$ corresponds to the identity time-warping $\gamma_{\rm id}$ and thus $E(x) = 0$.

\subsubsection{Model for functional combined principal components} \label{sec: FCPCA }

To model the association between $y$ and $x$, we define a random function $g^{C}$ on the extended domain $[0,2]$ for a positive scaling parameter $C>0$,
\begin{equation}
g^{C}(t) =
\begin{cases}
y(t), & t \in [0,1),\\
Cx(t-1), & t \in [1,2].
\end{cases}
\label{equ:6}
\end{equation}
The exclusion of the end point $\{1\}$ of the domain $[0,1]$ of $y$ in the construction of $g^{C}$ does not lose any information since $y$ is assumed to be continuous. Note that for any $y,x,C$, we have $g^{C} \in L_2[0,2]$. The parameter $C$ is introduced to adjust \emph{scaling imbalance} between $y$ and $x$. We will discuss the role of $C$ shortly, but for now we let $C$ be fixed.

For a given $C$, denote the eigen-decomposition of the covariance function $\Sigma_{g^{C}}$ of $g^{C}$ by
\begin{align*}
\Sigma_{g^{C}}(s,t) = \sum_{i=0}^{\infty} \lambda_{i}^{C} \xi_{i}^{C}(s) \xi_{i}^{C}(t), \quad s,t \in [0,2],
\end{align*}
where $\lambda_{i}^{C}$ are eigenvalues of $\Sigma_{g^{C}}$ in the decreasing order ($\lambda_{i}^{C} \ge \lambda_{i+1}^{C} \ge 0$, $i \ge 1$), and $\xi_{i}^{C}$ is the eigenfunction corresponding to $\lambda_i^C$.
The eigenfunctions are orthonormal, i.e., $\| \xi_{i}^{C} \|_2 = 1$ and $\langle \xi_{i}^{C}, \xi_{j}^{C} \rangle = 0$ for $i \neq j$.
The superscript $C$ is used to emphasize the dependence of the decomposition on $C$.
By Karhunen-Lo\'eve decomposition, we write $g^{C}(t) = \mu(t) + \sum_{i=1}^{\infty} z_{i}^{C} \xi_{i}^{C}(t)$,
$t \in [0,2]$, where $z_i^C$'s are uncorrelated mean-zero random variables with $E((z_{i}^{C})^2)=\lambda_{i}^{C}$. Note that the mean function $\mu=E(g^{C})$ does not depend on $C$ since $y$ is irrelevant of $C$ and $E(x)=0$. The function $g^C$ is then divided into the amplitude and phase functions as
\begin{equation}
\begin{split}
y^{C}(t) &= \mu(t) + \sum_{i=1}^{\infty} z_{i}^{C} \xi_{i}^{C}(t),\quad t \in [0,1),\\
x^{C}(t) &= \sum_{i=1}^{\infty} \frac{z_{i}^{C}}{C} \xi_{i}^{C}(t+1),\quad t \in [0,1].
\end{split}
\label{equ:8}
\end{equation}
In~(\ref{equ:8}), the joint variation between $y$ and $x$ is paired in eigenfunctions $\xi_{i}^{C}$.

The role of the scaling parameter $C$ in (\ref{equ:6}) becomes clear from (\ref{equ:8}). As opposed to the unit-free $x$, values of $y$ depend on the unit in which measurements of $y$ (or $f$) are made. The overall analysis should not depend on the particular scaling of $y$ (due to, for example, changes from the metric system to US customary units). Since scaling of $y$ by $C$ is equivalent to scaling of $x$ by $C^{-1}$, we introduce the scaling parameter $C$ applied only to the ``$x$ part'' of $g^{C}$, in order to keep the original unit of observed $f$ (and $y$). The eigenfunctions $\{ \xi_{i}^{C} \}_{i=1}^{\infty}$ and their eigenvalues $\{ \lambda_{i}^{C} \}_{i=1}^{\infty}$  vary for different choices of $C$; for a small $C$, the first few eigenfunctions $\xi_{i}^{C}$ are bound to capture more variations from the amplitude variation, while for a large $C$, the leading eigenfunctions reflect more phase variations. For any given $f$, or the pair $(y,x)$, there exists a continuum of different sets $\{\xi_{i}^{C}\}_{i=1}^{\infty}$, depending on the value of $C \in (0, \infty)$, which causes an identifiability issue. To our aim of succinctly representing the combined variation of $y$ and $x$ in the original function space, we choose $C$ to be dependent on the original random function $f$ as discussed below.

Let $m$ be a positive integer. From (\ref{equ:5}) and (\ref{equ:8}), for a given $C>0$, we define $A_{m}^{C}(f)$ as a \emph{projection} of $f$ onto the $m$-dimensional eigen-space, spanned by the first $m$ eigenfunctions, by
\begin{equation}
A_{m}^{C}(f)(t) = y^{C}_{m} \left (\int_0^t \text{Exp}_{\mu}^2(x^{C}_{m})(s) ds \right ),\quad t \in [0,1),
\label{equ:9}
\end{equation}
where for $t \in [0,1)$,
\begin{equation}
\begin{split}
y^{C}_{m}(f)(t) &= \mu(t) + \sum_{i=1}^{m} z_{i}^{C} \xi_{i}^{C}(t), \\
x^{C}_{m}(f)(t) &= \sum_{i=1}^{m} \frac{z_{i}^{C}}{C} \xi_{i}^{C}(t+1).
\end{split}
\label{equ:10}
\end{equation}
This projection utilizes the standard orthogonal projection of $g_C$ to its eigen-space in $L_2[0,2]$, but is \emph{non-linear} in the original function space $L_2[0,1]$. To minimize the approximation error of $A_{m}^{C}(f)$ with respect to $f$, the scaling parameter $C: = C_m$ is chosen as follows:
\begin{equation}
\label{equ:18}
C = \text{argmin}_{C >0} E\left[d^{2}(A_{m}^{C}(f), A_{\infty}^{C}(f))\right] = \text{argmin}_{C >0} E\left[d^{2}(A_{m}^{C}(f), f)\right],
\end{equation}
where $d$ is a distance function on $L_2[0,1]$. We use $d(f,g) = \|f-g\|_2$ for fast computation and mathematical convenience. Other distance functions such as  $L_1$-distance,  Fisher-Rao distance \citep{13}, and the earth-mover's distance \citep{40} can be used as well.

For a chosen $C$, the combined principal component of $y$ and $x$ (or the so-called eigen-mode) can be visualized in the original function space. In particular, the $i$th \emph{mode of variation} of $f$ can be visualized by overlaying the curves $\tilde{f}_{i,z} := \tilde{y}_{i,z} \circ \phi^{-1}(\tilde{x}_{i,z})$ for various values of $z \in \Re$.
Here, $\tilde{y}_{i,z}$ and $\tilde{x}_{i,z}$ are obtained from (\ref{equ:8}) by setting $z_{i}^{C}=  z \sqrt{\lambda_{i}^{C}}$ and also setting $z_{j}^{C}=0$ for all $j \not= i$.
Figure \ref{fig:1}(d) shows empirical estimates of $\tilde{f}_{1,z}$, $z=-1,0,1$, for the Berkeley data.
Our estimation procedure is described in Section \ref{sec:estimation}.

We note that one may use approaches of multivariate functional principal component analysis \citep[cf.][]{chiou2014multivariate,Happ2016} instead of gluing the two functions as done in (\ref{equ:6}). While such multivariate approaches may be mathematically more appealing, using (\ref{equ:6}) facilitates our discussion for the adaptive choice of $C$, and is satisfactory in our numerical examples.

\subsubsection{Model for correlation analysis} \label{sec: FCCCA }

As another approach to model the association between $y$ and $x$, we briefly discuss a model for a functional \textit{combined} canonical correlation analysis (CCA).

For a pair of non-random functions $\psi_y, \psi_x \in L_2[0,1]$, write $\rho(\psi_{y},\psi_{x})$ for the correlation coefficient between two random variables $\langle \psi_{y},y \rangle$ and $\langle \psi_{x}, x \rangle$. Here, $y$ and $x$ are the random amplitude and phase functions as defined before.
In functional combined CCA, the association between the amplitude  and  phase functions is modeled by a few canonical weight function pairs $(\psi_y,\psi_x)$ that sequentially maximize $\rho(\psi_{y},\psi_{x})$.
 In general, the $i$th canonical weight function pair $(\psi_{y,i}, \psi_{x,i})$ maximizes $\rho(\psi_{y,i},\psi_{x,i})$, with the constraint that
$\mbox{Cov}(\langle \psi_{y,i}, y \rangle, \langle \psi_{y,j}, y \rangle)
 = \mbox{Cov}(\langle \psi_{x,i}, x \rangle, \langle \psi_{x,j}, x \rangle)=0$ for $1 \le j < i$. The correlation coefficient $\rho_{i} := \rho(\psi_{y,i},\psi_{x,i})$ is called the $i$th canonical correlation coefficient.

The joint variation modeled by the $i$th canonical weight functions $\psi_{y,i}$ and $\psi_{x,i}$ can be visualized in the original function space. For $a,b \in \Re$, let
\begin{equation}
\begin{split}
&P_{y,(i,a)}(t) = \mu(t) + a \psi_{y,i},\quad t \in [0,1],\\
&P_{x,(i,b)}(t) = b \psi_{x,i},\quad t \in [0,1].
\end{split}
\label{equ:13}
\end{equation}
Then the $i$th mode of variation given by the functional combined CCA is visualized by overlaying the curves of $\tilde{f}_{i,a,b}:=  P_{y,(i,a)} \circ \phi^{-1}( P_{x,(i,b)}) $ for various values of $(a,b)$. A reasonable choice of $(a,b)$ satisfies $a/b = \beta$, where $\beta$ is the slope from the regression of $\langle \psi_{xi}, x \rangle$ against $\langle \psi_{yi},y \rangle$
%
%

\section{Estimation} \label{sec:estimation}

In this section we discuss our procedures for the application of functional combined PCA and CCA to a data set.


\subsection{Decomposition into amplitude and phase functions} \label{sec:3.1decomposition_empirical}

Let $f_{i}$, $i = 1,\ldots,n$, be the $i$th realization of the underlying random function $f$ obtained from $n$ independent experiments. The realizations $f_{i}$'s do not manifest themselves in a direct way. They are usually recorded at discrete time points, leading to observed values $f_{ij}$, at time point index $j = 1,\ldots,n_i$, and sometimes are blurred with measurement errors. We assume that smoothing the observations $\{ f_{ij} \}_{j=1}^{n_{i}}$ with a suitable basis function system gives a close approximation of $f_{i}$. Denote the approximations to $f_i$ by $\hat{f}_i$, $i = 1,\ldots,n$.

Each $\hat{f}_i$ is then decomposed into the amplitude and phase functions by applying the method of Fisher-Rao function alignment \citep{13} to all sample $\{ \hat{f}_{i} \}_{i=1}^{n}$, which iteratively time-warps $\hat{f}_i$ to a template function, resulting in the time-warp $\hat\gamma_i$ and the aligned function $\hat{y}_i$, satisfying
\begin{equation}
\hat{f}_{i}(t) = \hat{y}_{i}(\hat{\gamma}_{i}(t)), \,\, i=1,2,..,n, \,\, t \in [0,1].
\label{equ:15}
\end{equation}
Write $\hat{x}_i = \phi(\hat\gamma_i)$. The Fisher-Rao alignment is known to be invariant to the choice of templates, and we choose it to satisfy $\sum_{i=1}^n \hat{x}_i = 0$ so that the center of $\{ \hat\gamma_i \}_{i=1}^{n}$ is $\gamma_{\rm id}$. Other methods of function alignment may be used here. We use the method of \citet{13} for its good performance \citep{14,44} and invariance to the choice of templates.

\subsection{Functional combined PCA} \label{sec:FCPCAestimation}

In the model for the functional combined PCA, the population eigen-structure depends on the unknown parameter $C$. We first discuss the empirical eigen-decomposition for any given $C$, and then present our procedure to obtain a data-adaptive estimate of $C$.

\subsubsection{Estimation of $(\mu, \lambda_{i}^{C}, \xi_{i}^{C})$}

Let the scaling parameter $C$ be given. For easy computation, we evaluate the functions $\hat{y}_{i}$ and $\hat{x}_{i}$ on a fine grid, $0= t_1 <t_2< \cdots<t_k = 1$, to obtain their vector expressions $\hat{\mathbf{y}}_{i}$ and $ \hat{\mathbf{x}}_{i}$. Write
\begin{align*}
\hat{\mathbf{g}}_{i}^{C} =
\begin{bmatrix}
\hat{\mathbf{y}}_{i}\\
C\hat{\mathbf{x}}_{i}
\end{bmatrix},\,\,\,
\hat{\mathbf{y}}_{i}=[\hat{y}_{i}(t_1)  \,\, \dots \,\, \hat{y}_{i}(t_k)]^{T}, \,\, \hat{\mathbf{x}}_{i}=[\hat{x}_{i}(t_1)  \,\, \dots \,\,\hat{x}_{i}(t_k)]^{T},
\end{align*}
and let $\hat{\boldsymbol{\mu}}=\sum_{i=1}^{n} \hat{\mathbf{g}}_{i}^{C} / n$.
The eigen-decomposition of the sample covariance matrix $\widehat{\Sigma}_{g^{C}}$ obtained from $\{ \hat{\mathbf{g}}_{i}^{C} \}_{i=1}^{n}$ provides $(n-1)$ pairs of eigenvalues and eigenvectors $(\hat{\lambda}_{i}^{C}, \hat{\boldsymbol{\xi}}_{i}^{C})$,
\begin{align*}
\widehat{\Sigma}_{g^{C}}=\sum_{i=1}^{n} [\hat{\mathbf{g}}_{i}^{C}-\hat{\boldsymbol{\mu}}][\hat{\mathbf{g}}_{i}-\hat{\boldsymbol{\mu}}]^{T}=\sum_{i=1}^{n-1} \hat{\lambda}_{i}^{C} \hat{\boldsymbol{\xi}}_{i}^{C} \left(\hat{\boldsymbol{\xi}}_{i}^{C}\right)^{T},
\end{align*}
where $\hat{\lambda}_{1}^{C}  \ge \dots \ge \hat{\lambda}_{n-1}^{C} \ge 0$, $\| \hat{\boldsymbol{\xi}}_{i}^{C} \|_2=1$ and $\langle \hat{\boldsymbol{\xi}}_{i}^{C}, \hat{\boldsymbol{\xi}}_{j}^{C} \rangle=0$ for $i \not= j$.
Estimates $\hat{\mu}$ of $\mu$ and $\hat{\xi}_{i}^{C}$ of $\xi_{i}^{C}$ are obtained by interpolation of the elements of $\hat{\boldsymbol{\mu}}$ and $\hat{\boldsymbol{\xi}}_{i}^{C}$.

\subsubsection{Estimation of $C$}

The estimates $\{ (\hat{\lambda}_{i}^{C}, \hat{\xi}_{i}^{C}) \}_{i=1}^{n-1}$ are dependent on the value of $C$. We note that the true parameter $C$ depends on the number of principal components, $m$, used in (\ref{equ:18}). For the purpose of exploratory analysis and visualization of the data, $m$ is typically chosen as a small number. For a given $m$, our strategy in the estimation of $C$ is to use an empirical minimizer of (\ref{equ:18}).
For this, let $a_{ij}^{C}=\langle \hat{g}_{i}^{C} - \hat\mu, \hat{\xi}_{j}^{C} \rangle$ be the $j$th score of the $i$th observation. We write $A_{m}^{C}(\hat{f}_{i})$ for an approximation of the $i$th observation $\hat{f}_i$ by the first $m$ empirical principal components, which is defined by (\ref{equ:9}), by replacing $y^{C}_{m}$ and $x^{C}_{m}$ with
\begin{align*}
\hat{y}^{C}_{m}(\hat{f}_{i})(t) &= \hat{\mu}(t) + \sum_{j=1}^{m} a_{ij}^{C} \hat{\xi}_{j}^{C}(t), \,\, t \in [0,1),\\
\hat{x}^{C}_{m}(\hat{f}_{i})(t) &= \sum_{j=1}^{m} \frac{a_{ij}^{C}}{C} \hat{\xi}_{j}^{C}(t+1), \,\, t \in [0,1].
\end{align*}
Our choice of $\hat{C}$ is then
\begin{equation}
\hat{C} = \underset{C >0}{\text{argmin}} \sum_{i=1}^{n} \frac{\| {A}_{m}^{C}(\hat{f}_{i}) - \hat{f}_{i} \|_2^{2}}{n},
\label{equ:16}
\end{equation}
which entails that the first $m$ combined principal components $\hat{\xi}_{i}^{\hat{C}}$ found at $C=\hat{C}$ reconstruct $\{ \hat{f}_{i} \}_{i=1}^{n}$ most faithfully, compared to other values of $C$. In practice, we use a numerical method to solve (\ref{equ:16}), which is almost instantaneous for small values of $m$.

In all of our numerical studies, the minimizer $\hat{C}$ always exists, and does not degenerate to 0 nor diverges to infinite. Heuristically, this is because we assume that the observation has both amplitude and phase variations. Large (or small) values of $C$ force the eigenfunctions $\hat{\xi}_{i}^{C}$ to explain only the phase variation (or amplitude variation, respectively). For large $C$, the amplitude variation of $\hat{f}_i$ is typically not found in $\hat{A}_m^C(\hat{f}_i)$; for small $C$, the two functions $\hat{f}_i$ and $\hat{A}_m^C(\hat{f}_i)$ exhibit different phases.
%
%
%

\subsection{Functional combined CCA}

In the functional combined CCA of the data $\{f_i: i =1,\ldots,n\}$, we again use the decomposed functions $(\hat{y}_i, \hat{x}_i)$, obtained in Section \ref{sec:3.1decomposition_empirical}, to compute estimates of the triple $(\rho_{j}, \psi_{y,j}, \psi_{x,j})$ as defined in Section \ref{sec: FCCCA }.

It is well know that a naive adaptation of the usual CCA procedure to functional data often leads to spurious estimates of the triple with the estimated canonical correlation coefficient close to one. Following the suggestions in \citet{19}, we use the regularized functional CCA as follows. For a given smoothing parameter $\lambda>0$, the estimates of the canonical weight functions are
 \begin{equation}
(\hat{\psi}_{y,1}, \hat{\psi}_{x,1}) = \max_{\psi_{y},\psi_{x} \in L_2[0,1]} \widehat{\mbox{Cov}}(\langle \psi_{y},\hat{y}_{i} \rangle, \langle \psi_{x}, \hat{x}_{i} \rangle)
\label{equ:17}
\end{equation}
subject to $\widehat{\mbox{Var}}(\langle \psi_{y},\hat{y}_{i} \rangle) + \lambda \| D^2\psi_{y} \|_2^2 = \widehat{\mbox{Var}}(\langle \psi_{x}, \hat{x}_{i} \rangle) + \lambda \| D^2\psi_{y} \|_2^2 = 1$, where $\widehat{\mbox{Cov}}$ and $\widehat{\mbox{Var}}$ denote sample covariance and variance and $D^2$ is the second order differential operator. Subsequent pairs $(\hat{\psi}_{y,j}, \hat{\psi}_{x,j})$ are obtained similarly with the additional orthogonality constraint. The $i$th empirical canonical correlation coefficient $\hat\rho_{i}$ is given by the sample correlation coefficient of $(\langle\psi_{y,1},\hat{y}_i\rangle,\langle\psi_{x,1},\hat{x}_i\rangle)$. We refer to \citet{1} for a detailed procedure of the functional CCA and the choice of $\lambda$ by a generalized cross-validation.

\section{Combined analysis of amplitude and phase variations in real data sets} \label{sec:real data}

In this section, we illustrate applications of the proposed methods to two sets of real data.

\subsection{Berkeley growth data} \label{sec:Berkeley}

The Berkeley growth data set \citep{6} consists of the height measurements of 39 boys and 54 girls from age 1 to 18. We present here the results of our analysis for the boy-only data. The analysis for girls' growth leads to a similar conclusion. To highlight periods of slower and faster growths, we use the growth velocity curves, by taking derivatives of the smoothed growth curves. These raw data are shown in Fig.~\ref{fig:1}(a).

The application of the proposed functional combined PCA and CCA results in a succinct dimension reduction of the data, as well as interpretable major modes of variations.  In particular, the first two combined principal components (PCs) well explain the association between  growth velocities (amplitude variation) and temporal paces (phase variations).

\begin{figure}[t!]
\centering
\includegraphics[width=1\textwidth]{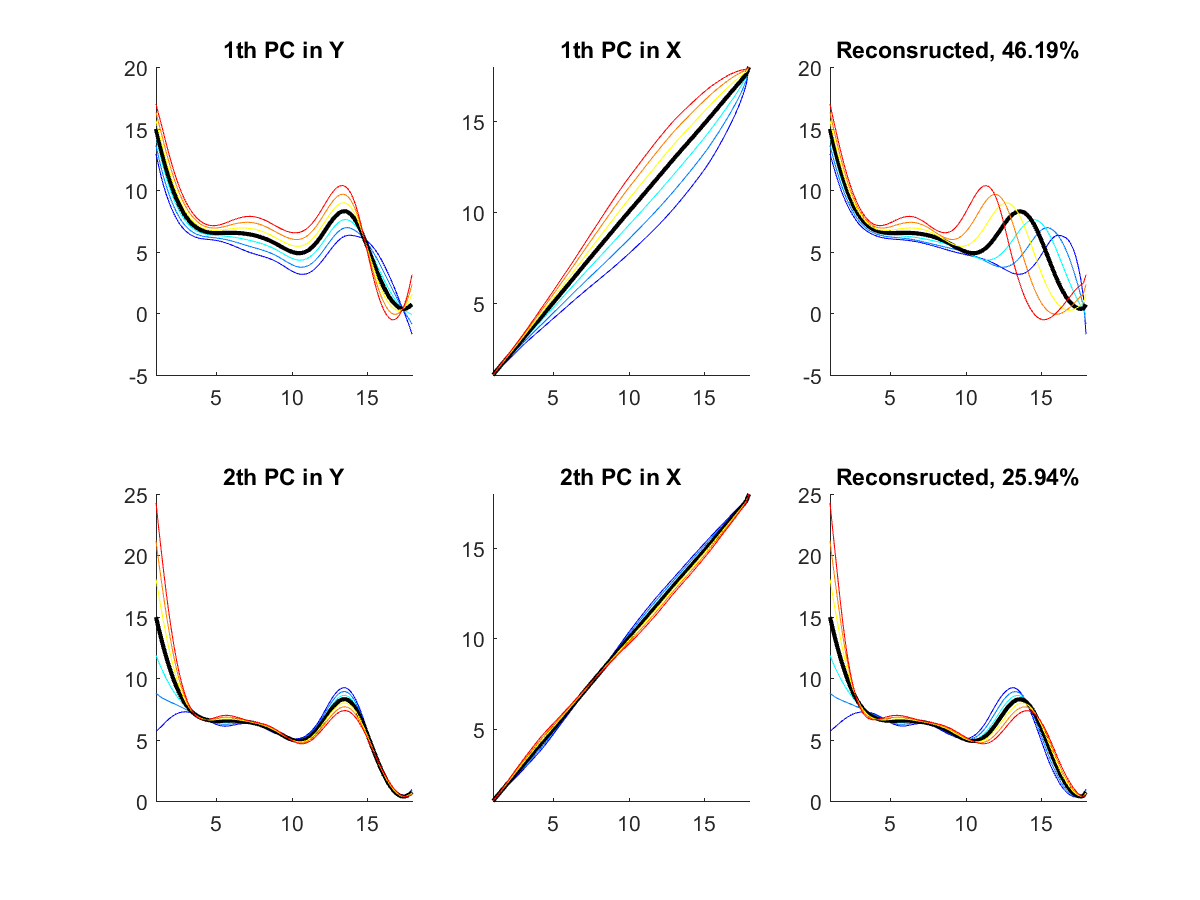}
\caption{First two combined principal component scores from the growth data. Amplitude functions (left column) and phase functions (middle column) are combined in the right column. Colors correspond across columns.}
\label{fig:X and Y}
\end{figure}

The mode of variation captured in the first combined PC explains the pattern that boys with higher overall growth rates tend to have fast temporal paces (e.g., reaching their pubertal growth spurt earlier than others). In the first row of Fig.~\ref{fig:X and Y}, the red curves represent this patten. On the other hand, boys with lower growth rates tend to have slower paces, as shown in the figure by the blue curves.
The second combined PC (shown in the second row of Fig.~\ref{fig:X and Y}) captures a contrast, which is characterized by the growths before and after about 9 years old. Specifically, the second PC explains a growth pattern that the growth rate and pace are positively associated for growths in ages 0--9, and negatively associated for growths in ages 10--18. As mentioned earlier, FCPCA aims to simultaneously capture the amplitude, phase and their association, and does so for this data set. The interpretable modes of variation shown in Fig.~\ref{fig:X and Y} are not typically found in applications of functional PCA (see e.g. Fig.~\ref{fig:1}).

\begin{figure}[t!]
\centering
\includegraphics[width=1\textwidth]{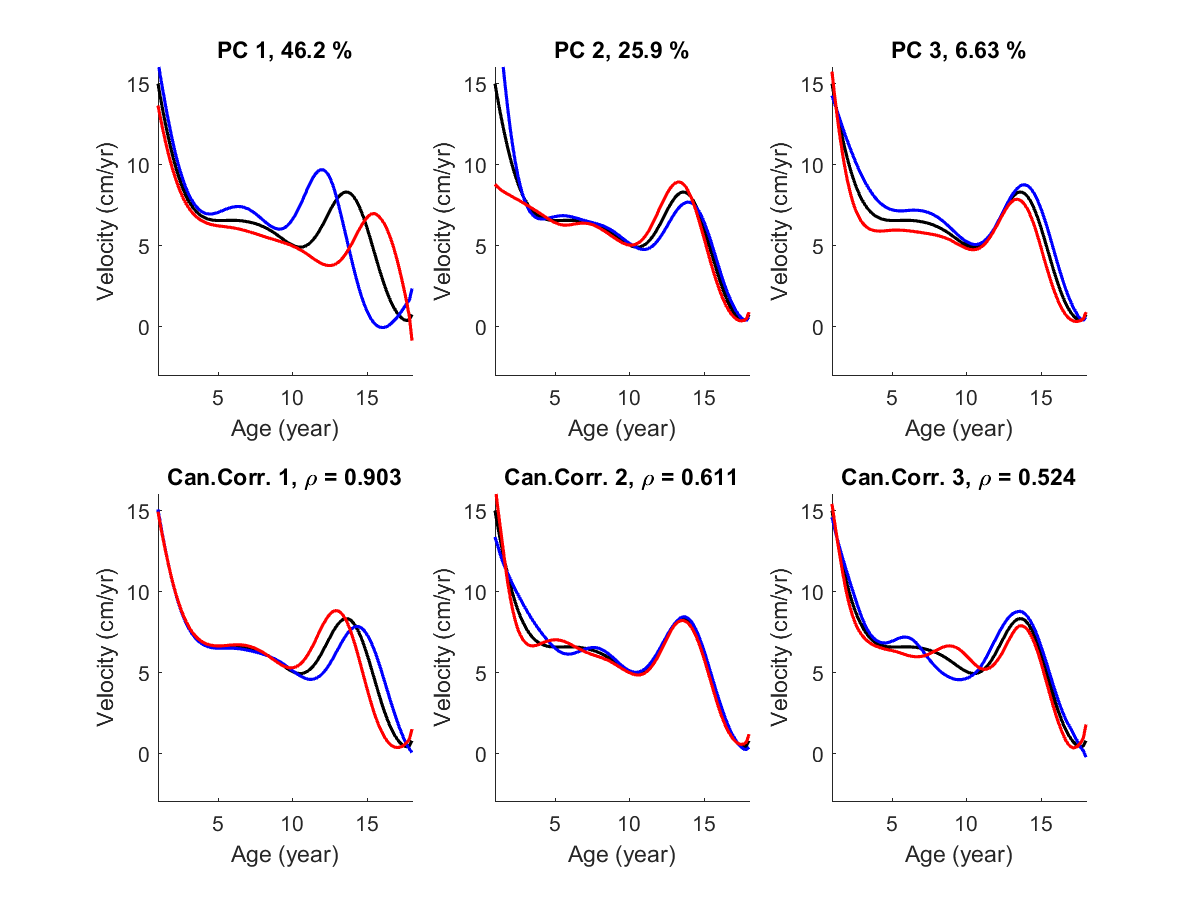}
\caption{Berkeley growth data. Top row: Major modes of variations captured in the functional combined PCA. Bottom row: Major associations between the amplitude and phase from the functional combined CCA.}
\label{fig:7}
\end{figure}

An application of our functional combined CCA to the data set reveals a difference between two of our proposed methods. The reconstructed functions from the most correlated components are shown in the botton row of Fig.~\ref{fig:7}.
These are visually different from the combined principal components shown in the top row.
The differences in patterns found by functional combined PCA and CCA should not be surprising. The \emph{internal} variations within each of amplitude and phase functions affect the combined PCA, while, in CCA, they are simply ignored.

\subsection{Lip motion data} \label{sec:Lip}

\begin{figure}[t!]
\centering
\includegraphics[width=1\textwidth]{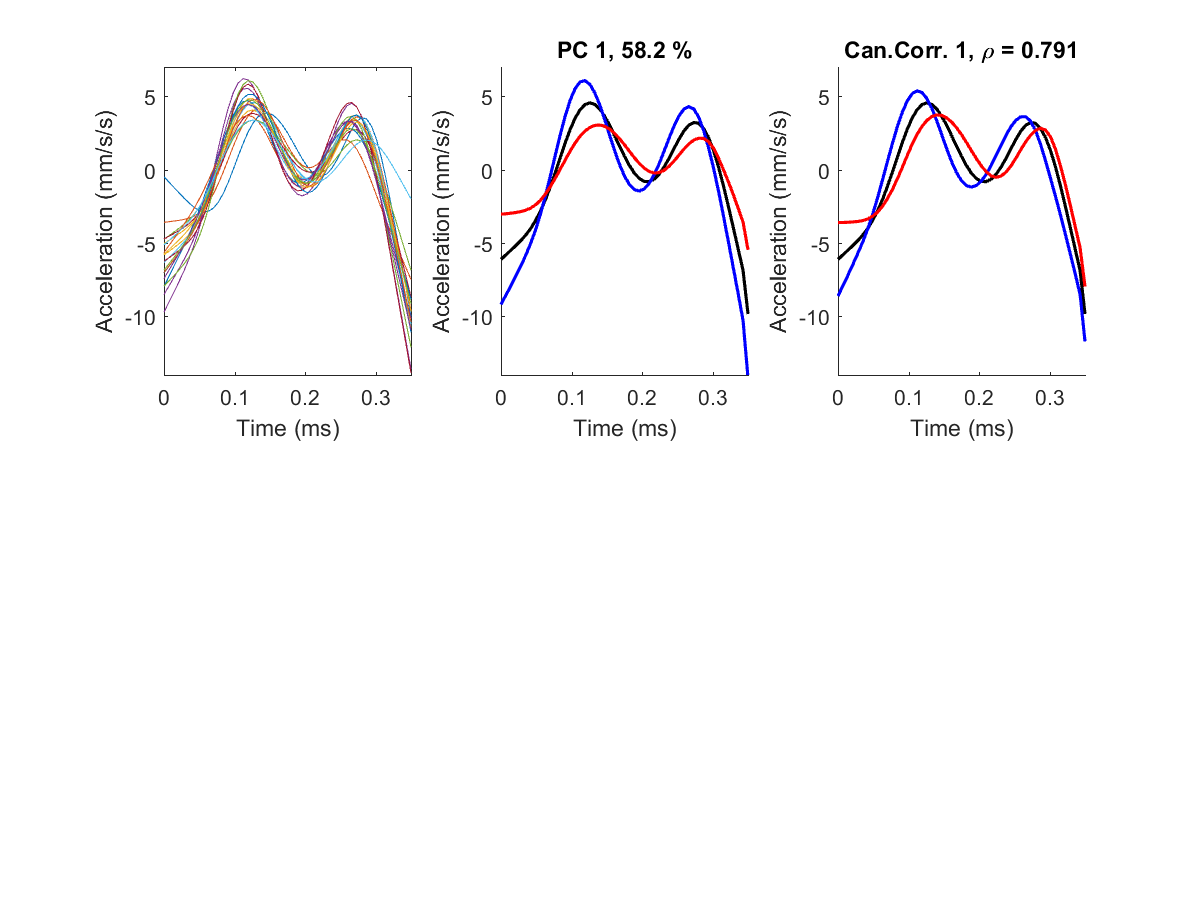}
\vspace{-2.5in}
\caption{(a) 20 acceleration curves of lip movement. (b) Three functions describing a first mode of variation from FCPCA. (c) Three functions describing a combined effect of the most correlated directions from the functional combined CCA.}

\label{fig:8}
\end{figure}

The data set we analyze here is a part of lip motion data used in \cite{17}. The data set is composed of measurements at 51 equally-spaced points in the timeframe from 0 to 340 milliseconds of a vertical position of lower lip while the subject speaks a syllable ``bob'' 20 times. The dynamics of lip motion is well captured by its acceleration. These second derivatives plotted in Figure \ref{fig:8}(a) show a common pattern. Lip movement is first accelerated negatively and then pass through a positive acceleration phase during which the descent of the lower lip is stopped. This lip opening phase is followed by a short period of near zero acceleration when pronunciation of the vowel ``o'' is at its full force, followed by another strong acceleration upward initiating lip closure. The movement is  completed by a negative acceleration episode as the lip returns to the closed position \citep{17}.

By an application of FCPCA, we found that the first combined PC explains a large portion (58\%) of the total variation. The first mode of variation, shown in Figure \ref{fig:8}(b), explains a speech habit of the speaker; as he makes the sound of the word louder (or softer), he tends to speak faster (or slower, respectively). For this data set, the findings from the function combined CCA are similar to those of FCPCA.

\section{Numerical studies} \label{sec:numerical studies}


\subsection{Efficiency of functional combined PCA under non-linear associations} \label{sec:non-linear}

The success of the proposed methods depends on whether a particular type  of the association between the amplitude and phase variations exists in data. In particular, our methods are well-suited for a linearly dependent amplitude $y$ and phase $x$ functions. To elaborate this point, we present a toy data analysis.

\begin{figure}[t!]
\centering
\includegraphics[width=1\textwidth]{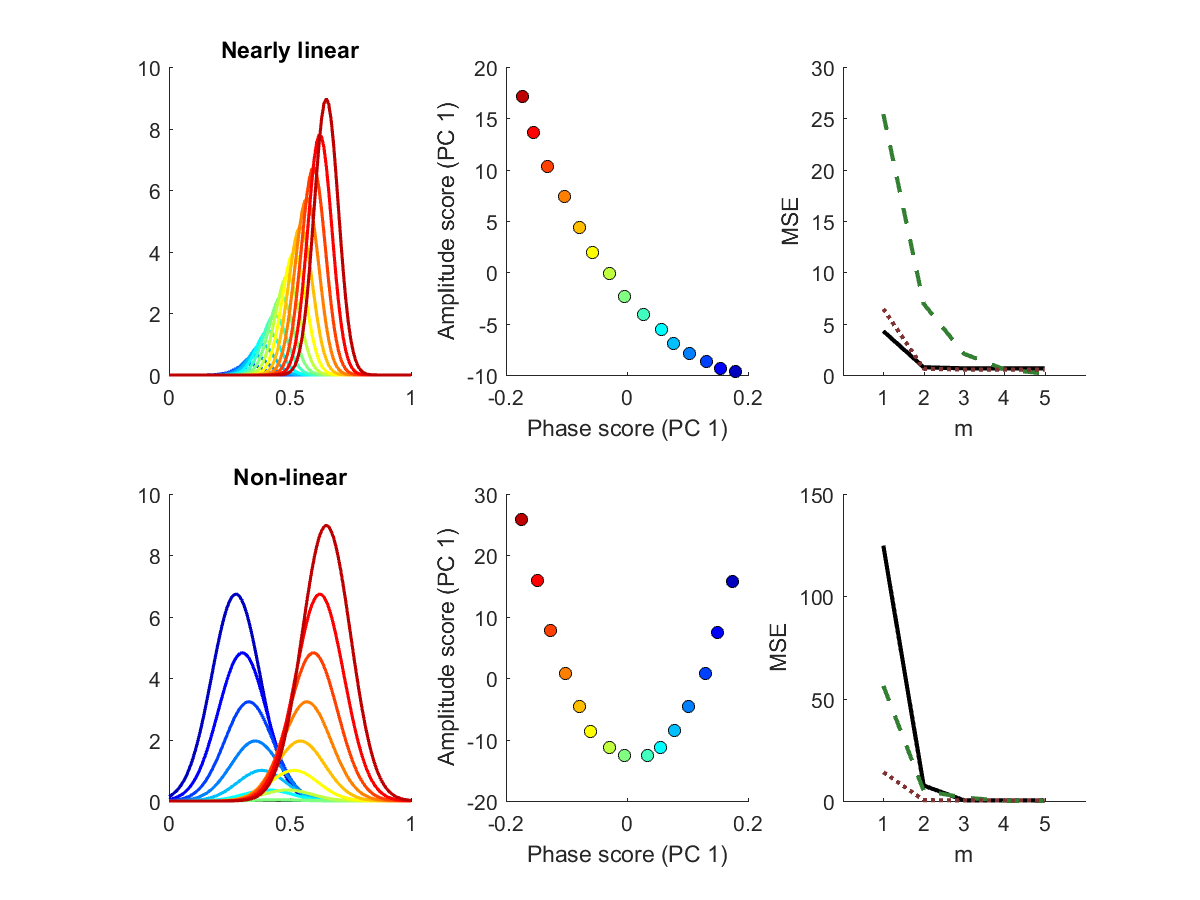}
\caption{Reconstruction errors of functional combined PCA (black solid), FPCA (green dashed) and the composite method (red dotted) of \citet{16}. The proposed method works well when the amplitude and phase are \emph{linearly} associated.}
\label{fig:6}
\end{figure}

Two sets of data are prepared by sampling from the amplitude and phase function pair $(y,x)$. We have set each of $y$ and $x$ has one major principal component, and the association between the PC score of $y$ and that of $x$ is either nearly linear or severely non-linear (quadratic). The observations are obtained by the composition, $f = y \circ \phi^{-1}(x)$, and displayed in the first column of Fig.~\ref{fig:6}. The types of association, or the degrees of non-linearity, are illustrated in the scatters of the two individual PC scores, shown in the second column of Fig.~\ref{fig:6}. The proposed functional combined PCA works well for the first data set, where the association between $y$ and $x$ is nearly linear.

To confirm this and to investigate the sensitivity of our method to the degrees of non-linearity, we evaluate for each data set the mean squared approximation error (MSE) using only the first $m$ components, as a function of $m \ge 1$, computed by $n^{-1} \sum_{i=1}^{n}\| {A}^{\hat{C}}_m(\hat{f}_i) - \hat{f}_i \|_2^2$. These errors are compared with errors from other natural competitors: the usual functional PCA (FPCA) and a \emph{composite} functional PCA, proposed in \cite{16}. The FPCA is applied to the original data (without applying function alignment), and the first $m$ components are used to approximate the observations.
In the composite method, the FPCA is applied to each individual functions ($\hat{y}$ and $\hat{x}$). First $m$ components from both $\hat{y}$ and $\hat{x}$ are used to approximate the observations (thus using $2m$ components). These MSEs are shown in the last column of Fig.~\ref{fig:6}.

Note that our definition of one-dimensional linear or nearly linear association, as shown in the first row of Fig.~\ref{fig:6}, typically results in a one-dimensional non-linear mode of variation in the original function space. This non-linear variation is not completely captured in a single component of FPCA, and oftentimes needs multiple components.
In contrast, our method efficiently captures the non-linear variation (showing the smallest MSE for $m = 1$), since in fact, the non-linear association becomes linear in $\hat{y}$ and $\hat{x}$.
 For this type of association, the usual FPCA needs several components to capture the non-linear variation in the original space, and is less favorable. The separate method, on the other hand, uses $2m$ linear components (compared to only $m$ components in the other two methods), thus is expected to show better performances than FPCA in general. Note that our functional combined PCA has smaller errors than the separate method has for this data set.

As the degrees of non-linearity intensify, the advantage of the functional combined PCA gradually lessens. For the severely non-linear case (shown in the second row of the figure), our method fails to capture the non-linear mode of variation in one component. However, it performs  comparable to other methods when more than one component is used, i.e. for $m>1$.

\subsection{Performance of estimation in functional combined PCA}\label{sec:simPCA}

In this and next subsections, we exhibit good performances of our estimation procedures. The success of our methods is largely dependent upon the quality of the alignment. The Fisher-Rao function alignment we choose to use has been shown to work well in practice \citep{14}, but its theoretical results (e.g., consistency in the estimation of $\mu$) are limited \citep{13}. Instead, we use simulated data sets to glimpse the consistency of the estimators. We have tried a range of parameter settings, and the results are concordant across settings. Below we present representative cases.

\begin{figure}[t!]
\centering
\includegraphics[width=1\textwidth]{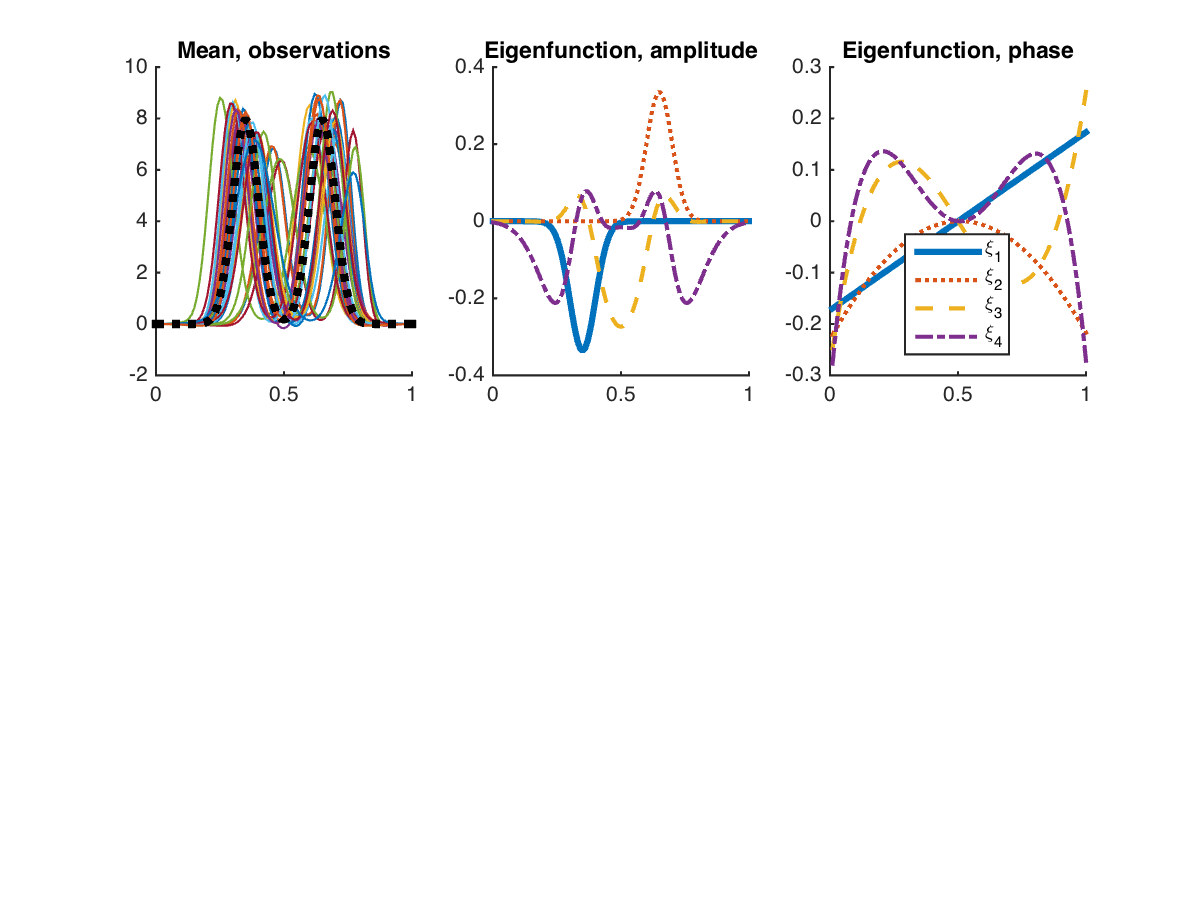}
\vspace{-2.7in}
\caption{Model used in Sections~\ref{sec:simPCA} and  \ref{sec:simCCA}. }
\label{fig:sim_model1}
\end{figure}

We use a four-component model for (\ref{equ:6}), where $g_{i}^{C}(t) = \mu(t) + \sum_{j=1}^{4} z_{ij} \sqrt{\lambda_j}  \xi_{j}(t)$, $t \in [0,2]$.
We set $\mu(t) = 20 [ g((t - 0.35)/0.05) + g((t-0.65)/0.05)]$, $t \in [0,1)$, where $g(\cdot)$ is the density function of the standard normal. The eigenfunctions for amplitudes are chosen by the Gram-Schmidt orthogonalization of four functions
$$
\left\{  g\left(\frac{t-0.35}{0.05}\right), g\left(\frac{t-0.65}{0.05}\right), g\left( \frac{t-0.5}{0.1}\right), g\left(\frac{t-0.3}{0.1}\right) + g\left(\frac{t-0.7}{0.1}\right) \right\},
$$
while the eigenfunctions for phases are from
$ \{(t - 0.5)^j : j = 1,\ldots, 4\}$.  Figure~\ref{fig:sim_model1} illustrates the mean function, and the eigenfunctions. We set $C = 1$ and
$(\lambda_1,\ldots, \lambda_4) = (3.5,2.6,0.3,0.1)$. The scores $z_{ij}$ are sampled from the standard normal distribution. The observed function $f_i$ is obtained from $g_i^C$ using (\ref{equ:5}) and (\ref{equ:8}). An example of such random sample is shown in the left panel of Fig.~\ref{fig:sim_model1}.

We observe $f_i$ at each time point $t_j: = (j-1)/ 101$, for $j=1,2,\ldots,101$ with measurement error $\epsilon_{ij} \sim N(0,0.1)$. As for a smoothing step for $f_{ij}$'s, the B-spline basis system of degree 4 with a roughness penalty on second derivative is used. Following~\cite{20}, knots are placed at evaluation points $\{ t_k \}_{k=1}^{101}$ and, following~\cite{21}, the value of the smoothing parameter $\lambda$ is determined by the generalized cross-validation method.


%

\begin{figure}[t!]
\centering
\includegraphics[width=1\textwidth]{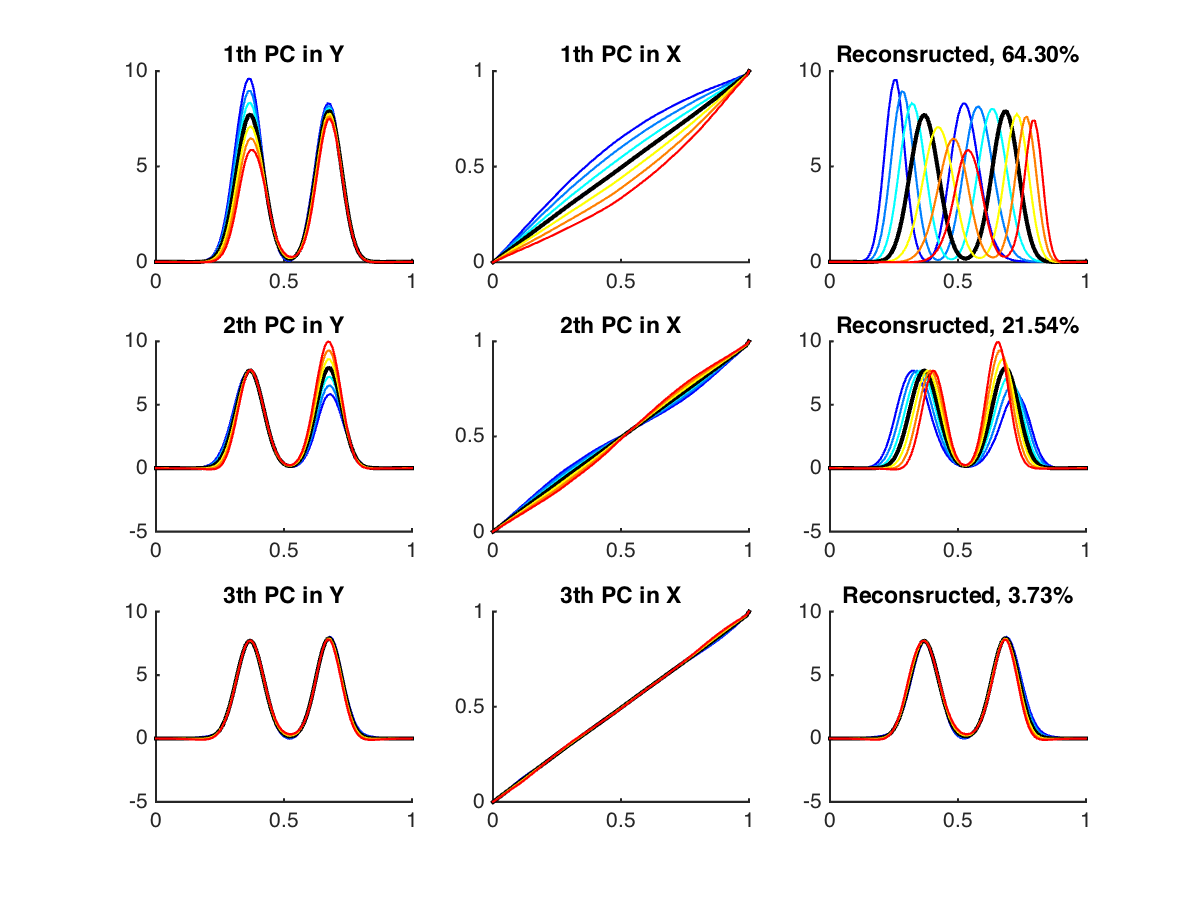}
\caption{Modes of variations captured in FCPCA for simulated data in Section~\ref{sec:simPCA}.}
\label{fig:sim_result}
\end{figure}

For sample sizes $n = 30,100$, we generated $f_1,\ldots,f_n$ from the model described above and obtained the estimates $(\hat{C}, \hat{\mu}, \hat{\lambda}_{1}^{\hat{C}}, \hat{\xi}_{1}^{\hat{C}}, \hat{\lambda}_{2}^{\hat{C}}, \hat{\xi}_{2}^{\hat{C}})$, from our procedure discussed in Section~\ref{sec:FCPCAestimation}. For a random sample of size $n = 30$, the analysis result is shown in Fig.~\ref{fig:sim_result}. There, we see that the first two component estimates capture the amplitude, phase and their association rather well; the estimates are very close to the population eigenfunctions, shown in Fig.~\ref{fig:sim_model1}. The third component seems negligible as $\lambda_3$ is small.

We repeat the experiment 100 times to witness the sampling distributions of the estimators. The result is summarized in Table \ref{tab1}. We observed that the estimators approach their population counterparts as the sample size increases.

\begin{table}[t]
\centering
\renewcommand{\arraystretch}{1.4}
\begin{tabular}{c|cc}
               & $n = 30$ & $n = 100$ \\
\hline
$ \hat{C} $   ($C=1$)                 & 1.44 (0.31) & 1.28 (0.29)\\
$ \hat{\lambda}_{1}^{\hat{C}}$  ($\lambda_{1}=3.5$)& 4.12 (0.26)& 3.81 (0.21) \\
$ \hat{\lambda}_{2}^{\hat{C}}$  ($\lambda_{2}=2.6$)& 2.98 (0.37)& 2.74 (0.18) \\
$ \| \mu - \hat{\mu}\|_2                $&2.89 (1.27)&2.15 (0.85)\\
$ \|\xi_{1}-\hat{\xi}_{1}^{\hat{C}}\|_2 $&0.49 (0.24)&0.38 (0.36)\\
$ \|\xi_{1}-\hat{\xi}_{2}^{\hat{C}}\|_2 $&0.71 (0.41)&0.34 (0.50)\\
\end{tabular}
\caption{Simulation results for functional combined PCA. The mean and standard deviation (in parentheses) of scalar estimates and $L_2$-distances of functional estimates to their parameter counterparts are shown for different sample sizes.
\label{tab1}}
\end{table}

\subsection{Performance of estimation in functional combined CCA}\label{sec:simCCA}

For a model for the functional combined CCA, the amplitude and phase functions are each modeled using four principal components, where
$y_{i}(t) = \mu_{y}(t) + \sum_{i=1}^{4} u_{i} \sqrt{\lambda_{y,i}} \xi_{y,i}(t)$,
$x_{i}(t) = \sum_{j=1}^{4} v_{j} \sqrt{\lambda_{x,i}} \xi_{x,j}(t)$, $t \in [0,1]$,
so that the corresponding function $f$ is obtained by the function composition (\ref{equ:5}).
We use $\mu$, $\xi_{y,i}$, $\xi_{x,j}$ as shown in Fig.~\ref{fig:sim_model1}.
We choose to model only one canonical weight function pair by setting $(\psi_{y,1}, \psi_{x,1}) := (\xi_{y,1}, \xi_{x,2})$ with the canonical correlation coefficient 0.8. (That is, only the first ``y'' component and the second ``x'' component are correlated.)
The variances of individual principal components are set to be $(\lambda_{y,1},\ldots, \lambda_{y,4}) = (5,3.5,0.8,0.7)$, and $(\lambda_{x,1},\ldots,\lambda_{x,4}) = (1,0.7,0.16,0.14) / 100$. The scores $(u_i,v_j)$ are independently sampled from $N(0,1)$, except that $\mbox{Cov}(u_1,v_2) = 0.8$.
The random function $f_i$ is observed at a dense grid with a measurement error drawn from $N(0, 0.1)$ and the data are processed as done in Section~\ref{sec:simCCA}.

We obtained the empirical sampling distributions of the estimators $(\hat{\rho}_{1}, \hat{\psi}_{y,1}, \hat{\psi}_{x,1})$ for sample sizes $n =30,100$ with 100 repetitions. The results, summarized in Table \ref{tab2}, suggest a good performance of our estimation procedure. Note that we have used the generalized cross validation \citep{1} to choose the smoothing parameter of functional CCA. 

\begin{table}[t]
\centering
\renewcommand{\arraystretch}{1.4}
\begin{tabular}{c|cc}
               & $n = 30$ & $n = 100$ \\
\hline
$ \hat{\rho}_1 $   ($\rho_1=0.8$)                 &0.68 (0.21)&0.72 (0.19)\\
$ \|\psi_{y,1}-\hat{\psi}_{y,1}\|_2              $&0.89 (0.31)&0.43 (0.18) \\
$ \|\psi_{x,1}-\hat{\psi}_{x,1}\|_2              $&0.72 (0.28)&0.55 (0.15)\\
\end{tabular}
\caption{Simulation results for functional combined CCA. The mean and standard deviation (in parentheses) of $\hat\rho$ and $L_2$-distances of functional estimates to their parameter counterparts are shown for different sample sizes.
\label{tab2}}
\end{table}

\section{Conclusion}

This paper presents a novel framework for exploring the combined structure of amplitude and phase variations in functional data. Naive applications of standard statistical tools such as the functional PCA to this type of data sometimes produces unsatisfactory results. The commonly-employed framework of statistical analysis of aligned functions by the use of function alignment disregards the relevant phase variation. To overcome the disadvantages, we propose functional combined PCA and CCA to investigate major modes of variation and correlated directions of data in the underlying space, in which the association between amplitude and phase variations can be addressed. The analysis results are visually presented in the original form of observed functions to aid interpretation.

 %
%
%
%
%
%

\bibliographystyle{Chicago}
\bibliography{mybibfile}

\begin{thebibliography}{}

\bibitem[\protect\citeauthoryear{Chen and M{\"u}ller}{Chen and
  M{\"u}ller}{2012}]{chen2012nonlinear}
Chen, D. and H.-G. M{\"u}ller (2012).
\newblock Nonlinear manifold representations for functional data.
\newblock {\em The Annals of Statistics\/}~{\em 40\/}(1), 1--29.

\bibitem[\protect\citeauthoryear{Chiou, Chen, and Yang}{Chiou
  et~al.}{2014}]{chiou2014multivariate}
Chiou, J.-M., Y.-T. Chen, and Y.-F. Yang (2014).
\newblock Multivariate functional principal component analysis: A normalization
  approach.
\newblock {\em Statistica Sinica\/}~{\em 24\/}(4), 1571--1596.

\bibitem[\protect\citeauthoryear{Craven and Wahba}{Craven and Wahba}{1979}]{21}
Craven, P. and G.~Wahba (1979).
\newblock Smoothing noisy data with spline functions.
\newblock {\em Numerische Mathematik\/}~{\em 31\/}(4), 377--403.

\bibitem[\protect\citeauthoryear{de~Boor}{de~Boor}{2001}]{20}
de~Boor, C. (2001).
\newblock {\em A Practical Guide to Splines}.
\newblock Springer.

\bibitem[\protect\citeauthoryear{Gasser, M{\"u}ller, K{\"o}hler, Molinari, and
  Prader}{Gasser et~al.}{1984}]{3}
Gasser, T., H.-G. M{\"u}ller, W.~K{\"o}hler, L.~Molinari, and A.~Prader (1984).
\newblock Nonparametric regression analysis of growth curves.
\newblock {\em The Annals of Statistics\/}~{\em 12\/}(1), 210--229.

\bibitem[\protect\citeauthoryear{Gervini}{Gervini}{2015}]{46}
Gervini, D. (2015).
\newblock Warped functional regression.
\newblock {\em Biometrika\/}~{\em 102\/}(1), 1--14.

\bibitem[\protect\citeauthoryear{Hadjipantelis, Aston, M{\"u}ller, and
  Evans}{Hadjipantelis et~al.}{2015}]{47}
Hadjipantelis, P., J.~Aston, H.-G. M{\"u}ller, and J.~Evans (2015).
\newblock Unifying amplitude and phase analysis: A compositional data approach
  to functional multivariate mixed-effects modeling of mandarin chinese.
\newblock {\em Journal of the American Statistical Association\/}~{\em
  110\/}(510), 545--559.

\bibitem[\protect\citeauthoryear{Hadjipantelis, Aston, M{\"u}ller, and
  Moriarty}{Hadjipantelis et~al.}{2014}]{48}
Hadjipantelis, P., J.~Aston, H.-G. M{\"u}ller, and J.~Moriarty (2014).
\newblock Analysis of spike train data: A multivariate mixed effects model for
  phase and amplitude.
\newblock {\em Electronic Journal of Statistics\/}~{\em 8\/}(2), 1797--1807.

\bibitem[\protect\citeauthoryear{Happ and Greven}{Happ and
  Greven}{2016}]{Happ2016}
Happ, C. and S.~Greven (2016).
\newblock Multivariate functional principal component analysis for data
  observed on different (dimensional) domains.
\newblock {\em Journal of the American Statistical Association\/}~(to appear).

\bibitem[\protect\citeauthoryear{Jung, Dryden, and Marron}{Jung
  et~al.}{2012}]{JungPNS}
Jung, S., I.~L. Dryden, and J.~S. Marron (2012).
\newblock {Analysis of Principal Nested Spheres}.
\newblock {\em Biometrika\/}~{\em 99\/}(3), 551--568.

\bibitem[\protect\citeauthoryear{Karcher}{Karcher}{1977}]{karcher1977riemannian}
Karcher, H. (1977).
\newblock Riemannian center of mass and mollifier smoothing.
\newblock {\em Communications on pure and applied mathematics\/}~{\em 30\/}(5),
  509--541.

\bibitem[\protect\citeauthoryear{Kneip and Ramsay}{Kneip and Ramsay}{2008}]{9}
Kneip, A. and J.~Ramsay (2008).
\newblock Combining registration and fitting for functional models.
\newblock {\em Journal of the American Statistical Association\/}~{\em
  103\/}(483), 1155--1165.

\bibitem[\protect\citeauthoryear{Kurtek, Wu, Christensen, and
  Srivastava}{Kurtek et~al.}{2013}]{14}
Kurtek, S., W.~Wu, G.~Christensen, and A.~Srivastava (2013).
\newblock Segmentation, alignment and statistical analysis of biosignals with
  application to disease classification.
\newblock {\em Journal of Applied Statistics\/}~{\em 40\/}(6), 1270--1288.

\bibitem[\protect\citeauthoryear{Leurgans, Moyeed, and Silverman}{Leurgans
  et~al.}{1993}]{19}
Leurgans, S., R.~Moyeed, and B.~Silverman (1993).
\newblock Canonical correlation analysis when the data are curves.
\newblock {\em Journal of the Royal Statistical Society\/}~{\em 55\/}(3),
  725--740.

\bibitem[\protect\citeauthoryear{Mardia and Jupp}{Mardia and
  Jupp}{2000}]{Mardia2000}
Mardia, K.~V. and P.~E. Jupp (2000).
\newblock {\em {Directional Statistics}}.
\newblock Wiley.

\bibitem[\protect\citeauthoryear{Marron, Ramsay, Sangalli, Srivastava,
  et~al.}{Marron et~al.}{2015}]{44}
Marron, J.~S., J.~O. Ramsay, L.~M. Sangalli, A.~Srivastava, et~al. (2015).
\newblock Functional data analysis of amplitude and phase variation.
\newblock {\em Statistical Science\/}~{\em 30\/}(4), 468--484.

\bibitem[\protect\citeauthoryear{Park and Ahn}{Park and
  Ahn}{2017}]{park2016clustering}
Park, J. and J.~Ahn (2017).
\newblock Clustering multivariate functional data with phase variation.
\newblock {\em Biometrics\/}~{\em 73\/}(1), 324--333.

\bibitem[\protect\citeauthoryear{Ramsay, K.G.~Munhall, and Ostry}{Ramsay
  et~al.}{1996}]{17}
Ramsay, J., V.~G. K.G.~Munhall, and D.~Ostry (1996).
\newblock Functional data analyses of lip motion.
\newblock {\em Acoustical Society of America\/}~{\em 99\/}(6), 3718--3727.

\bibitem[\protect\citeauthoryear{Ramsay and Silverman}{Ramsay and
  Silverman}{2005}]{1}
Ramsay, J. and B.~Silverman (2005).
\newblock {\em Functional Data Analysis\/} (Second ed.).
\newblock Springer.

\bibitem[\protect\citeauthoryear{R.D.Tuddenham and Snyder}{R.D.Tuddenham and
  Snyder}{1954}]{6}
R.D.Tuddenham and M.~Snyder (1954).
\newblock Physical growth of california boys and girls from birth to eighteen
  years.
\newblock {\em Univ. of Calif. Publications in Child Development\/}~{\em
  1\/}(2), 183--364.

\bibitem[\protect\citeauthoryear{Rubner, Tomasi, and Guibas}{Rubner
  et~al.}{2000}]{40}
Rubner, Y., C.~Tomasi, and L.~J. Guibas (2000).
\newblock The earth mover's distance as a metric for image retrieval.
\newblock {\em International journal of computer vision\/}~{\em 40\/}(2),
  99--121.

\bibitem[\protect\citeauthoryear{Sangalli, Secchi, Vantini, and
  Vitelli}{Sangalli et~al.}{2010}]{15}
Sangalli, L., P.~Secchi, S.~Vantini, and V.~Vitelli (2010).
\newblock K-mean alignment for curve clustering.
\newblock {\em Computational Statistics and Data Analysis\/}~{\em 54\/}(5),
  1219--1233.

\bibitem[\protect\citeauthoryear{Srivastava, Jermyn, and Joshi}{Srivastava
  et~al.}{2007}]{srivastava2007riemannian}
Srivastava, A., I.~Jermyn, and S.~Joshi (2007).
\newblock Riemannian analysis of probability density functions with
  applications in vision.
\newblock In {\em 2007 IEEE Conference on Computer Vision and Pattern
  Recognition}, pp.\  1--8. IEEE.

\bibitem[\protect\citeauthoryear{Srivastava, Wu, Kurtek, Klassen, and
  Marron}{Srivastava et~al.}{2011}]{13}
Srivastava, A., W.~Wu, S.~Kurtek, E.~Klassen, and J.~Marron (2011).
\newblock Registration of functional data using {Fisher-Rao} metric.
\newblock {\em arXiv preprint arXiv:1103.3817\/}.

\bibitem[\protect\citeauthoryear{Tucker, Wu, and Srivastava}{Tucker
  et~al.}{2013}]{16}
Tucker, J., W.~Wu, and A.~Srivastava (2013).
\newblock Generative models for functional data using phase and amplitude
  separation.
\newblock {\em Computational Statistics and Data Analysis\/}~{\em 60}, 50--66.

\bibitem[\protect\citeauthoryear{Yu, Lu, and Marron}{Yu et~al.}{2017}]{22}
Yu, Q., X.~Lu, and J.~Marron (2017).
\newblock Principal nested spheres for time-warped functional data analysis.
\newblock {\em Journal of Computational and Graphical Statistics\/}~{\em
  26\/}(1), 144--151.

\end{thebibliography}
\end{document}